\documentclass[12pt]{article}
\pdfoutput=1
    \newcommand{\be}[1]{\begin{equation}\label{#1}}
    \newcommand{\ba}[1]{\begin{eqnarray}\label{#1}}
    \newcommand{\ep}[1]{\epsilon_{#1}}

    \newcommand{\rd}{{\rm d}}

    \newcommand{\pa}[1]{\left(#1\right)}
    \newcommand{\paq}[1]{\left[#1\right]}

    \newcommand{\M}{{\rm M_{\rm P}}}

    \def\ee{\end{equation}}
    \def\ea{\end{eqnarray}}

\usepackage{graphicx,latexsym}
\usepackage{graphicx,epsf}
\usepackage{color}
\usepackage{epsfig}
\usepackage{amsmath}
\usepackage[T1]{fontenc}
\usepackage[utf8]{inputenc}
\usepackage{tikz}
\usepackage{pgfplots}
\usepackage{authblk}
\begin{document}
\title{Non-Canonical Inflation and Primordial Black Holes Production}
\author[1]{Alexander Y. Kamenshchik\thanks{Alexander.Kamenshchik@bo.infn.it}}
\author[2]{Alessandro Tronconi\thanks{Alessandro.Tronconi@bo.infn.it}}
\author[2]{Tereza Vardanyan\thanks{Tereza.Vardanyan@bo.infn.it}}
\author[2]{Giovanni Venturi\thanks{Giovanni.Venturi@bo.infn.it}}
\affil[1]{Dipartimento di Fisica e Astronomia and INFN, Via Irnerio 46,40126 Bologna,
Italy\\
L.D. Landau Institute for Theoretical Physics of the Russian
Academy of Sciences, Kosygin str. 2, 119334 Moscow, Russia}
\affil[2]{Dipartimento di Fisica e Astronomia and INFN, Via Irnerio 46,40126 Bologna,
Italy}
\date{}
\maketitle

\begin{abstract}
We study a mechanism for the amplification of the inflationary scalar perturbation when the inflaton field action is non-canonical, i.e. the inflaton kinetic term  has a non-standard form. For such a case the speed of sound of the perturbations generated during inflation is less than one and in general changes with time. Furthermore in such models, even when the scalar field potential is negligible, diverse inflationary attractors may exist. The possible effects of a speed of sound approaching zero during some stage of inflation may lead to a large amplification for the amplitude of the scalar spectrum which, on horizon re-entry during the radiation dominated phase, can collapse and form primordial black holes (PBH) of a mass $M_{\rm BH}\sim 10^{-15}M_{\odot}$ which may constitute a large fraction of the total Dark Matter (DM) today. 
\end{abstract}

\section{Introduction}
LIGO's direct observations of black holes (BHs) through the gravitational waves emitted during their merging has started a new exciting era for high energy physics and astronomy. Growing interest in BH physics rapidly led, among the various alternatives, to the proposal of reconsidering light BHs as candidates for much of the total DM in the Universe. In order to explain their abundance and mass spectra an intriguing possibility is that their seeds may have originated during the very early stages of our Universe and, in particular, in the inflationary era or during the subsequent reheating. Only later, during the radiation dominated era, the over-densities generated during inflation larger than some critical value may collapse and form BHs \cite{BHform}.  If this were the case more information about the physics at the Planck scale is expected in the next decade, besides  the data coming from CMB anisotropies and, eventually, the detection of primordial gravitational waves.\\
It is widely accepted that some dynamical mechanism has driven an inflationary phase \cite{inflation} at the very beginning of our Universe. The exact features of such a mechanism are still unclear but nowadays any viable hypothesis about the mechanism of inflation must satisfy a list of stringent constraints coming from the observational data. Many of such constraints derive from CMB observations \cite{Planck,pert} and are restricted to a quite narrow energy window of the primordial inflationary spectra. Within the slow-roll (SR) paradigm many models compatible with data have been put forward.\\
In the last few years the ideas behind the renormalization group approach, which have been successfully applied to many areas of physics at low, testable, energies,  have been employed in gravity and the physics of the early universe \cite{AS}. At the energy scale of inflation one should also consider different operators, which have no effect at long distance/low energy , since they may have an important role before the Big Bang. In particular, in this article we shall study the possible consequences of a modification of the kinetic term of a minimally coupled inflaton. Such a modification has relevant consequences on the inflationary evolution, as pointed out in \cite{kinf}, where the authors show that a nearly de Sitter stage can be achieved by simply modifying the inflaton kinetic term , without the need of any potential. Subsequently the corresponding evolution of inflationary perturbations was calculated, showing an amplification of the scalar spectrum w.r.t. the tensor one due to the negligible speed of sound $c_s$ during the  evolution. Later, observations severely ruled out such small values of $c_s$ since they lead to large non-gaussian features in the CMB incompatible with the observational constraints. Nonetheless the possibility of a non-canonical kinetic term for the inflaton, with a non-negligible potential, have been studied in the literature \cite{noncan} leading to a slow rolling phase driven by the potential and enhanced by the modified kinetic term. Such models can be made compatibile with observations.\\ 
In this article we shall restrict our attention to the possibile addition of a potential to the original k-inflation model proposed in \cite{kinf}. In the presence of a non-negligible potential, the system can have 2 quasi de Sitter attractors. The inflationary phase can then interpolate between them. When the potential is leading w.r.t. the modified kinetic term, inflation occurs close to the ``standard'' attractor and when potential becomes negligible w.r.t the kinetic term, inflation may continue close to the k-inflationary attractor. This twofold evolution may have important consequences since the latter phase may amplify the inflatonic fluctuations which are outside the CMB observational window and, if certain conditions are met, these over-densities may collapse to form PBHs. \\
The paper is organised as follows. In Sec. II the basic formulae are introduced and some dynamical features are illustrated. In the Sec. III the  phase of potential domination is studied and the mechanism of amplification is discussed. In the Section IV some toy models are presented and finally the conclusions are presented in Section V.

\section{Formalism}
We consider an homogeneous inflaton-gravity system with a generic kinetic term for the inflaton field. 
Let its total action be
\be{act}
S=\int d^4x\sqrt{-g}\paq{\frac{\M^2}{2}R+p(\phi,X)}
\ee
where $p(\phi,X)$ is the pressure of the scalar field, $X\equiv\frac{1}{2}g^{\mu\nu}\partial_\mu\phi\partial_\nu\phi$ and $g_{\mu\nu}$ is the metric tensor. The energy density associated with the inflaton field is $\rho=2X\partial p/\partial X-p$
and the equations governing the evolution of the system can be cast in the usual form:
\be{frkgeqs}
H^2=\frac{\rho}{3\M^2}\,,\;\dot \rho=-3H\pa{\rho+p}.
\ee
We consider the pressure and energy density which take the following forms
\be{klagV}
p=\frac{1}{2}k(\phi)\dot \phi^2+\frac{1}{4}L(\phi)\dot \phi^4-V(\phi),\;\rho=\frac{1}{2}k(\phi)\dot \phi^2+\frac{3}{4}L(\phi)\dot \phi^4+V(\phi).
\ee
The KG equation then is
\be{kVKG}
\ddot \phi+3H\dot \phi\frac{k+L\dot \phi^2}{k+3L\dot \phi^2}+\frac{V'+\dot \phi^2\pa{\frac{k'}{2}+\frac{3}{4}L'\dot\phi^2}}{k+3L\dot \phi^2}=0
\ee
where the speed of sound is given by
\be{csdef}
c_s^2\equiv \frac{\partial p/\partial X}{\partial p/\partial X+2 X \partial^2 p/\partial X^2}=\frac{k+L\dot \phi^2}{k+3L\dot \phi^2}.
\ee
For constant $V$ the system (\ref{frkgeqs}) has a de Sitter/flat space solution $\dot\phi=0$. On the other hand a non-trivial de Sitter solution with $\dot \phi^2=-\frac{k}{L}$ exists if $k$ and $L$ have opposite signs and
\be{dScond}
\frac{\rd}{\rd \phi}\pa{V+\frac{k^2}{4L}}=0\rightarrow V=-\frac{1}{4}\frac{k^2}{L}+\Lambda
\ee
with $\Lambda$ arbitrary constant. Correspondingly one has $\rho=\Lambda$.\\ 
A slight deformation of this last constraint could lead to a SR (nearly de Sitter) evolution which, in the limit for $k\rightarrow 1$ and $L\rightarrow 0$, does not reproduce the usual SR condition since the Eq. (\ref{dScond}) is peculiar to the non-canonical framework. On the other hand  a suitable potential  can also drive a SR phase.\\ 
Let us consider from here on $L(\phi)>0$ and $k=-1$ ($k$ can be set to $-1$ without loss of generality by a suitable field redefinition and provided $k(\phi)<0$).
If the condition (\ref{dScond}) is not satisfied exactly then
\be{SRcond0}
\frac{\rd}{\rd\phi}\pa{\frac{1}{L}-4V}=\epsilon f(\phi)
\ee
where $\epsilon$ is a small parameter. The exact solution $\dot \phi^2=\frac{1}{L}$ is then slightly modified and becomes $\dot \phi^2=\frac{1}{L}+\epsilon D(\phi)$. Then Eq. (\ref{kVKG}), to the first order in $\epsilon$, takes the following form
\be{kVKGep}
2\M D'+\sigma\sqrt{3}D=\frac{\M f}{2}
\ee
where $\sigma\equiv{\rm sgn}\paq{ H/\dot \phi}$ and, in particular, $\sigma\equiv{\rm sgn}\dot \phi$ when the Universe is expanding.
Eq. (\ref{kVKGep}) then has the following general solution
\be{solD}
D(\phi)=\exp\pa{-\sigma\frac{\sqrt{3}\phi}{2\M}}\paq{D_0+\int_{\phi_0}^\phi\exp\pa{\frac{\sigma\sqrt{3}	\bar\phi}{2\M}}\frac{f}{4}\rd\bar\phi}
\ee
where the decreasing contribution, proportional to the integration constant $D_0$, rapidly disappears. Such a contribution, in the presence of a non-negligible $f(\phi)$, describes the (transient) approach to the SR attractor. Correspondingly one can calculate the SR parameters $\ep{1}$ and $\ep{2}$. To the first order in $\epsilon$ we have
\be{ep12sr}
\ep{1}=6LD,\quad \ep{2}=\frac{\sqrt{3}}{3}\sigma \frac{f}{D}\M\pa{1-\frac{2}{3}\ep{1}-\sigma\frac{2\sqrt{3}}{\M}\frac{D}{f}}.
\ee
On integration by parts (\ref{solD}) and neglecting the non-rapidly decaying part of $D$ one finds
\be{Dibp}
D\simeq \frac{\sqrt{3}}{6}\sigma\M \paq{f(\phi)-\exp\pa{-\sigma\frac{\sqrt{3}\phi}{2\M}}\int_{\phi_0}^{\phi} \frac{\rd f}{\rd \bar \phi}.\exp\pa{\sigma\frac{\sqrt{3}\bar\phi}{2\M}}\rd\bar\phi}.
\ee
Then $\ep{2}\sim \mathcal{O}\pa{\ep{1}}\ll 1$ if the second contribution of the square bracket in Eq. (\ref{Dibp}) is negligible w.r.t. the non-decaying part of $D$ i.e. if
\be{srcondep2L}
\frac{2\sqrt{3}}{3}\frac{\M}{f}\frac{\rd f}{\rd \phi}\ll 1
\ee
which is a generalised SR condition for $f(\phi)$. The speed of sound is also small and to the first order is
\be{csSR0}
c_s^2\simeq\frac{\ep{1}}{12}.
\ee
We observe that a singularity at $\dot \phi^2=\frac{1}{3L}$ is  present in the modified KG equation. It can be removed by a suitable choice for the potential namely $V=1/\pa{12L}$. Otherwise one must choose a suitable form for $L$ and $V$ so as to not to cross the singularity.\\
If $f$ is negligible, $D$ only consists of the decaying part and one finds
\be{ep12trans}
\ep{1}=6L_0D_0\exp\pa{-\sigma\frac{\sqrt{3}\phi}{2\M}},\quad \ep{2}=-3
\ee
that is $\ep{1}$ is not slowly varying but still remains very close to zero and inflation continues until some other effect enters the dynamics.\\
Let us finally consider $L=L_0$ and $V=V_0\exp\pa{\alpha\phi/\M}$.  Eq. (\ref{kVKGep}) has an attractor which can be algebraically obtained from the ansatz $D=D_0 \exp\pa{\alpha\phi/\M}$ leading to 
\be{eqD0}
2\alpha D_0+\sigma\sqrt{3}D_0=-2\alpha V_0\Rightarrow D_0=\frac{-2\alpha V_0}{2\alpha+\sigma\sqrt{3}}
\ee
and correspondingly one finds
\be{SRVsmall}
\ep{1}=6DL,\quad \ep{2}=2\sqrt{3}\,\sigma\,\alpha.
\ee

\section{Potential Driven Slow Roll}
The SR evolution is the consequence of the smallness of the SR parameters $\ep{1}\equiv -\dot H/H^2$ and $\ep{2}\equiv \dot\epsilon_{1}/\pa{H\ep{1}}$. In the non-canonical context the smallness of $\ep{1}$ is equivalent to
\be{VSR}
G\ll L\dot\phi^2\; {\rm or}\; V/\dot\phi^2 \gg {\rm max}(1,L\dot\phi^2).
\ee 
with $G(\phi,\dot \phi)=L(\phi)\dot \phi^2-1$, and that of $\ep{2}$ leads to
\be{conep2NC}
\frac{\ddot \phi}{H\dot \phi}\ll1\;{\rm and}\;\frac{\dot G}{HG}\ll 1.
\ee
The addition of a potential to k-inflation can generate a slow rolling phase similar to the canonical one, provided the potential is leading w.r.t. the non-canonical kinetic term and this latter term is ``slowly'' varying in Hubble time units. 
Given (\ref{VSR}) and (\ref{conep2NC}) the modified KG equation can be approximated by
\be{SRKG}
3H\dot \phi \,G+V'\simeq 0\Rightarrow \dot \phi \,G\simeq -\frac{2\M}{\sqrt{3}}\frac{\rd \sqrt{V}}{\rd\phi}
\ee
and $H\simeq\sqrt{V/3\M^2}$. Moreover one finds
\be{conep1NCb}
\ep{1}\simeq \frac{\M^2}{2G}\pa{\frac{V'}{V}}^2\equiv \frac{\ep{1}^{(c)}}{G}
\ee
where $\ep{1}^{(c)}$ is the SR parameter in the canonical case.
The derivative of Eq. (\ref{SRKG}) w.r.t. time divided by $H\dot \phi \,G$ gives
\be{derKGSR2}
\frac{\ddot \phi}{H\dot \phi}+\frac{\dot G}{H G}\simeq -\frac{\M^2}{G}\pa{\frac{V''}{V}}+\ep{1}\equiv -\frac{\eta^{(c)}}{G}+\frac{\ep{1}^{(c)}}{G}.
\ee
Let us note that (\ref{conep1NCb}) and (\ref{derKGSR2}) are straightforward generalisations of the canonical SR relations for $\ep{1}$ and $\ddot\phi/(H\dot \phi)$ where the factor $G$ now appears ($G=1$ in the canonical case). The presence of $G$ can amplify or reduce the SR parameter w.r.t. the corresponding ones obtained for the same potential in the canonical framework.\\

\subsection{The Importance of Potential Driven SR (PDSR)}
K-inflation can lead to an inflationary phase close to $\dot \phi^2=1/L$ which is ruled out given the present observational constraints on the non-gaussian features ($f_{NL}$) of the temperature fluctuations in CMB. The Planck constraints on primordial NG in general single-field models of inflation provide the following most stringent bound \cite{Planck} on the inflaton sound speed:
\be{csplanck}
c_s\ge c_{s,\rm max}\quad (95\%\;{\rm CL})
\ee
where $c_{s,\rm max}=0.024$ in the case $\dot c_s=0$ and $c_{s,\rm max}$ a bit larger (and model dependent) for the case of $\dot c_s/(Hc_s)$ is of the order of the SR parameters.
The presence of a suitable potential may alleviate such serious problems, possibly giving rise to an inflationary expansion which is divided into two phases. During the early stages, the potential of the scalar field dominates and the modes which imprinted their features in the observable part of the CMB spectrum exit the horizon. For these modes the speed of sound is compatible with observations. In the second phase the field evolves close to attractor $\dot\phi^2\simeq 1/L$ and $c_s\ll 1$. The modes which exit the horizon during this latter phase are unconstrained by CMB observations. Their evolution is crucially determined by the speed of sound at horizon crossing which amplifies non-gaussianities (in a frequency window which cannot be tested by CMB) and the amplitude of the scalar perturbation as well. Through a mechanism which resembles that of ultra slow roll \cite{USR} inflation with an inflection point in the potential, for our model the amplitude of the scalar fluctuations can be so amplified that  the possibility of generating primordial black holes must be seriously considered.\\
Current CMB observations indicate that the amplitude of primordial curvature perturbations is about $10^{-9}$ (often called the COBE normalization) that of CMB observable scales. In order for the PBH formation after inflation to be effective enough to be observationally interesting, curvature perturbations for the scales involved in the process must be $\mathcal{O}(10^{-2} - 10^{-1})$, that is  at least 7 order of magnitude larger.
The amplitude of the scalar fluctuations when a generic sound speed is considered is given by
\be{Ps}
\mathcal{P}_s=\left.\frac{1}{8\pi^2\M^2}\frac{H^2}{\ep{1}c_s}\right|_{c_sk=a H}
\ee
with, for the case we are discussing, $c_s^2=G/\pa{3L\dot\phi^2-1}$. An amplification of $\mathcal{P}_s$ of 7 orders of magnitude requires extremely fine tuning within the framework of canonical inflation with an inflection point. If a non-canonical kinetic term is taken into account the amplification comes about simply as the result of the twofold inflationary phase. Let us note that the emergence of non-gaussian features, possibly enhanced by a small $c_s$, can modify the estimates regarding the amount of amplification needed for PBHs formation \cite{nogaus}. Such an enhancement has been estimated in the context of SR inflation with a non canonical kinetic term and has been proven to be large. To our knowledge the estimate of the amplification of non-gaussianities in the regime we are considering (without SR) has never been studied before. In \cite{Noller}, for example, the effects on non-gaussianities originated by the departure from both scale invariance and SR have been studied for a class of non canonical models. The authors show that a non trivial background evolution can mitigate the amplification expected for these models in the SR regime. To conclude, the possibility of the existence of a non-gaussian spectrum for the modes responsible for PBHs formation deserves caution and further studies. In the present analysis such a possible effect is neglected.\\ 
Close to de Sitter attractor $\dot \phi\simeq \sigma/\sqrt{L_0}$, one always has  $H\simeq 1/\pa{2\M\sqrt{3 L_0}}\equiv H_0$ and then $\phi(a)\sim 2\sqrt{3}\sigma \M\ln \pa{a/a_0}$. The resulting speed of sound evolves as $c_s=C_0a^{-\gamma_s/2}$ where $C_0$ is an integration constant and $\gamma_s$ is a model dependent parameter.  Eq. (\ref{Ps}) then takes the form
\be{Psnew}
{\mathcal P}_s=\frac{1}{96\pi^2C_0^3}\frac{H_0^2}{\M^2}\pa{\frac{C_0 k}{H_0}}^{\frac{3\gamma_s}{2+\gamma_s}}\propto \frac{H_0^{\frac{4-\gamma_s}{2+\gamma_s}}}{C_0^{\frac{6}{2+\gamma_s}}\M^2}k^{\frac{3\gamma_s}{2+\gamma_s}}
\ee
where a spectral index different from one is present ($n_s-1=\frac{3\gamma_s}{2+\gamma_s}$). If such an index is larger than one the scalar spectrum is amplified for modes which exit the horizon well after those imprinted in the CMB. One may roughly estimate  the number of e-folds, $\Delta N_{\rm amp}$, necessary in order to obtain a desired value of amplification through the ratio
\be{ratPs}
\frac{\mathcal{P}_{s,{\rm bh}}}{\mathcal{P}_{s,*}}=\frac{H_{\rm bh}^2}{H_*^2}\frac{\epsilon_{1,*}c_{s,*}}{12\,c_{s,{\rm bh}}^3}\sim \exp\pa{\frac{3}{2}\gamma_s\Delta N_{\rm amp}}.
\ee
Let us finally note that, if the speed of sound and $H$ vary significantly during inflation, the following relation holds
\be{deN}
\Delta N(M)\simeq 18.4-\frac{1}{12}\ln\pa{\frac{g_{*k}}{g_{*0}}}-\frac{1}{2}\ln\frac{M}{M_{\odot}}+\ln \frac{c_{s,{\rm bh}}}{c_{s,*}}+\ln\frac{H_*}{H_{\rm bh}}
\ee
where $\Delta N(M)$ is the number of e-folds between the horizon exit of CMB modes and that of the modes amplified and finally collapsing into PBHs with a mass $M$, $M_{\odot}$ is the solar mass, $g_{*k}$, $g_{*0}$ indicates the number of relativistic degrees of freedom at the BHs formation and at present. Finally $c_{s,*}$ and $H_*$ are evaluated at horizon exit for CMB modes and $c_{s,\rm bh}$ and $H_{\rm bh}$ must be evaluated when the spectrum reaches the critical threshold $\sim 10^{-2}$.\\
Let us note that the above result is qualitatively different from that generally obtained from an ultra SR phase which generates a flat spectrum and an amplification of curvature perturbations.
\begin{figure}[t!]
\centering
\epsfig{file=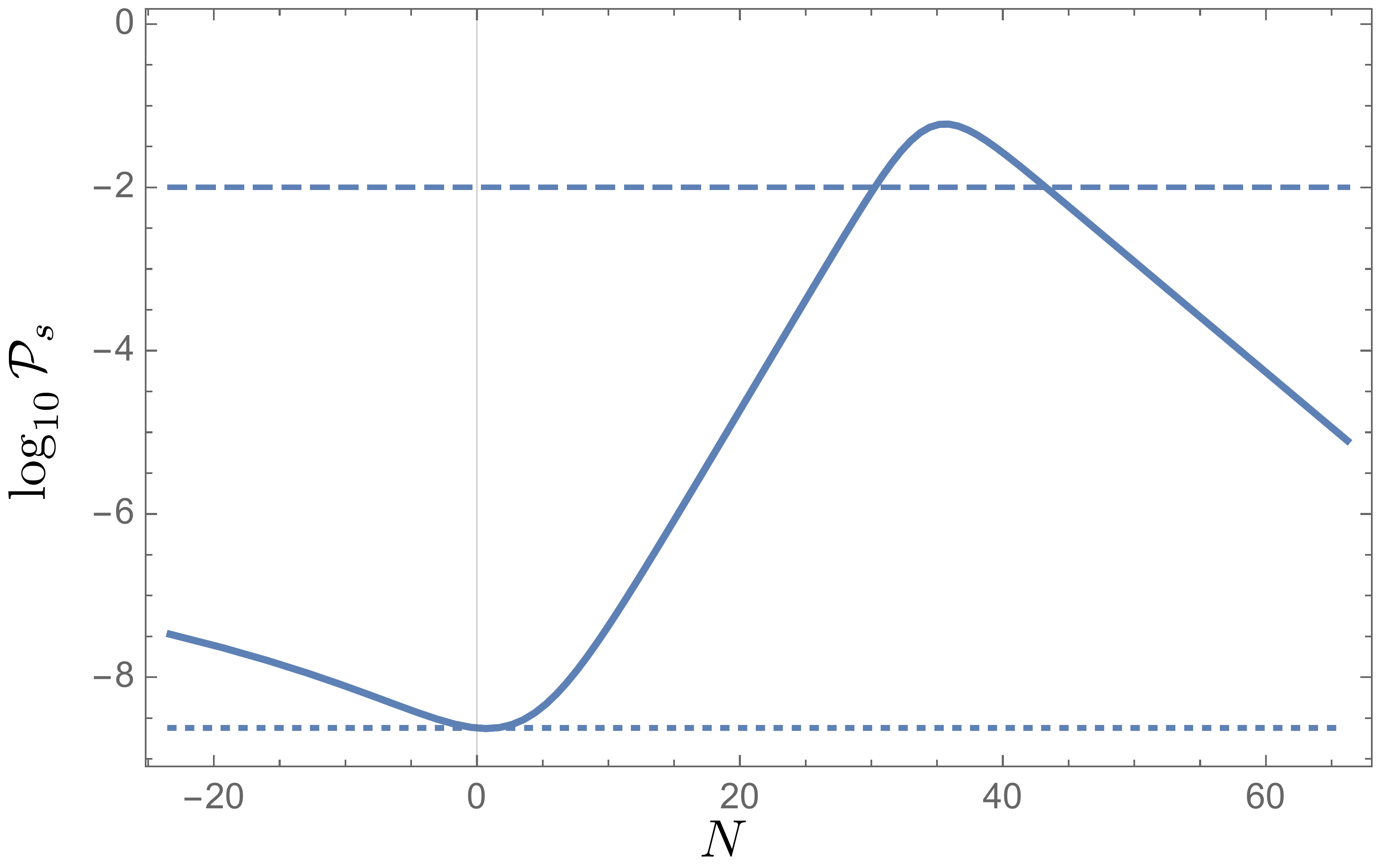, width=5.7 cm}
\epsfig{file=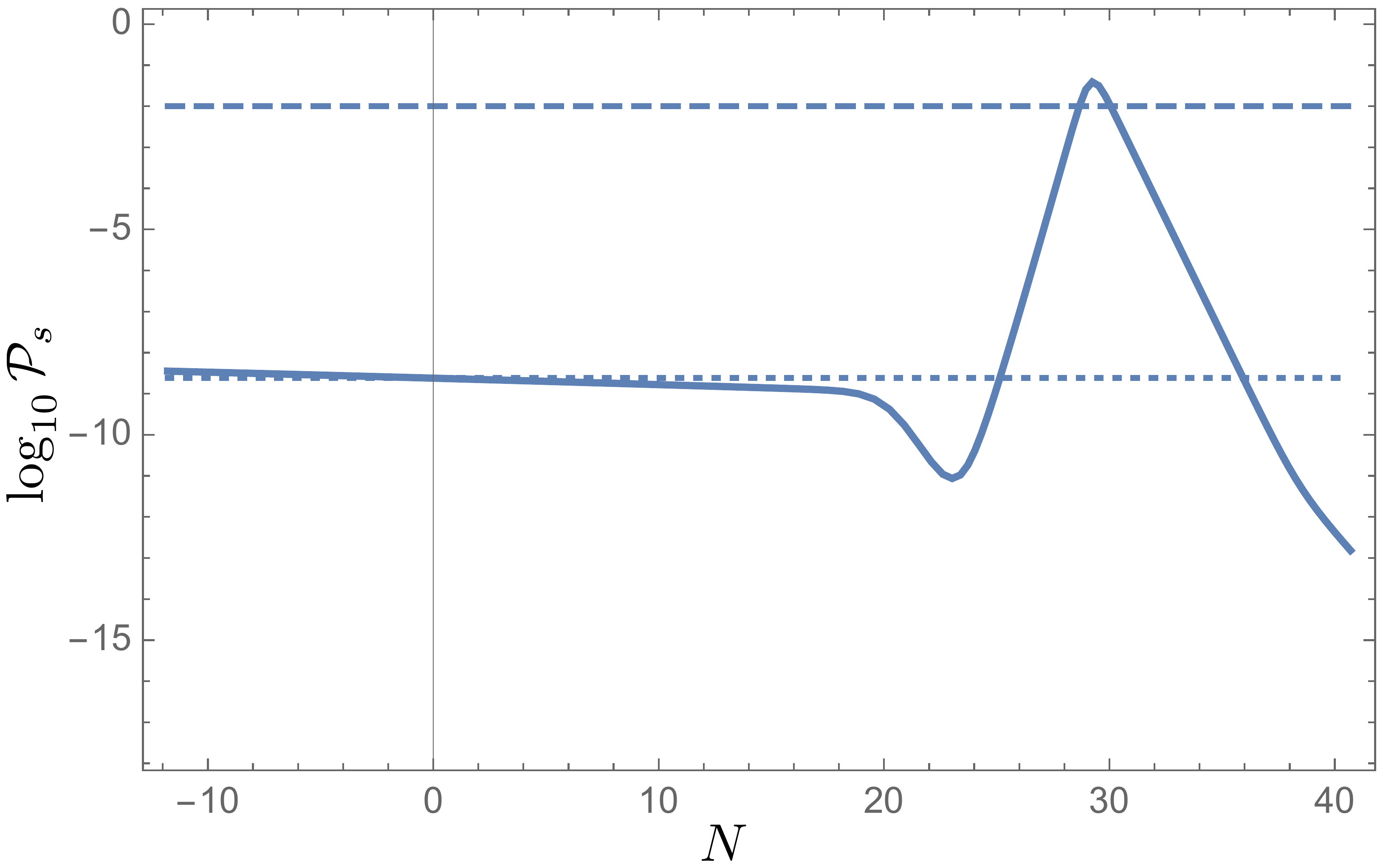, width=5.7 cm}
\caption{\it In the above figures the logarithm of the amplitude of the scalar spectrum is plotted as a function of $N$ for the potentials (\ref{pot1}) on the left and (\ref{pot2}) on the right. We set $N=0$ at the horizon exit of the CMB modes. The dotted line represents the observed amplitude of the scalar spectrum and the dashed line represents the critical value $10^{-2}$ for PBH formation.}
\label{f1}
\end{figure}
\section{Toy Models}
In order to illustrate how such an amplification may occur we consider the following toy models where the potential has a monotonic behaviour in order that the evolution is non-singular. Let us note that the models listed below are tuned in order to give $n_s\sim 0.965$ and ${\mathcal P}_s=2.4\cdot 10^{-9}$.\\
An exponential potential can accommodate many of the required features. Let us consider the following setup (Model I)
\be{pot1}
V=V_0\exp\pa{\frac{\alpha\phi}{\M}},\; L=L_0\paq{1+\delta \exp\pa{-\frac{ \alpha\phi}{2\M}}}
\ee
with $V_0=10^{-8}\M^4$, $\alpha\equiv\alpha_0=1.2\cdot 10^{-1}$, $L_0=1.15 \cdot 10^8\M^{-4}$ and $\delta=2\cdot 10^{-9}$.
For such a model $V_0$ and $L_0$ are tuned to reproduce the correct $n_s$ and ${\mathcal P}_s$. The value of $\delta$ and the exponential in $L$ are chosen in order to interrupt the amplification of ${\mathcal P}_s$ at the right time and not overproduce PBHs. Their form is not really important provided the variation of $L$ w.r.t $\phi$ is not too large compared to that of $V$. Let us note that the magnitude of $\alpha$ is crucial in determining both the magnitude of the slow roll parameters during PDSR and the amplification in the subsequent phase when the field approaches the attractor $\dot \phi\simeq \sigma/\sqrt{L}$. It can be related to variation of the speed of sound during the amplification phase and in particular $\gamma_s=2\alpha\sqrt{3}$. One then finds
\be{ampmod1}
10^7\simeq\frac{\mathcal{P}_{s,{\rm bh}}}{\mathcal{P}_{s,*}}\sim \exp\pa{3\alpha\sqrt{3}\Delta N}\Rightarrow \Delta N\sim 26
\ee
and correspondingly $M\sim 10^{-13}M_{\odot}$. The analytical approximation under-estimates $\Delta N$ which results in about $30$ e-folds by numerical simulations. Let us note that the $20\%$ deviation from the standard (ultra SR) case is due to the varying $H$ and $c_s$ is well predicted by analytical approximation. Let us further note that the analytical estimate (\ref{ratPs}) well approximates $\Delta N$ if CMB modes leave the horizon at $t_*$ which almost coincides with the beginning of the amplification stage. This is almost true for the potential (\ref{pot1}) (see Figure (\ref{f1})) but in general $\Delta N$ is the sum of two contributions $\Delta N\equiv \Delta N_I+\Delta N_{II}$ where $\Delta N_I$ is the e-fold interval between $t_*$ and the beginning of the amplification period while $\Delta N_{II}$ is the duration of this latter period.\\
Values of $\alpha$ smaller that $\alpha_0$ (see for example Model II with $\alpha=5\cdot 10^{-2}$ in table (\ref{tabexamples})) lead to a very flat spectrum and a very small tensor to scalar ratio, within the $68\%$ confidence level region estimated by the Planck data but correspondingly the required amplification of 7 orders of magnitude of ${\mathcal P}_s$  would need many e-folds ($\Delta N>40$). On the other hand larger $\alpha$'s would lead to a fast amplification, large values of the spectral indices and, thus, of the corresponding running and of the tensor to scalar ratio (see Model III for an example with $\alpha=3\cdot 10^{-1}$). In particular for this latter case an amplification of 8 orders of magnitude is needed in order to have a non-negligible production of PBHs.\\
We finally consider the following setup (Model IV) having a potential 
\be{pot2}
V=V_0\exp\pa{\frac{\alpha\phi}{15\M}}\paq{\tanh\pa{\frac{\alpha\phi}{\M}}+1},\; L=L_0\paq{1+\delta \exp\pa{-\frac{\alpha \phi}{\M}}}
\ee
with $V_0=1.85\cdot 10^{-10}\M^4$, $\alpha=5\cdot 10^{-1}$, $L_0=2.7 \cdot 10^9\,\M^{-4}$ and $\delta=2\cdot 10^{-12}$.
For such a case the first phase of inflation takes place for $\phi>0$ and a nearly flat potential with a small $\alpha_{\rm eff}\equiv \alpha/15$. This guarantees small SR parameters, running and tensor to scalar ratio. At $\phi<0$ the potential becomes negligible in (\ref{kVKGep}) and there is a fast approach to the attractor $\dot \phi=-L^{-1/2}$ described by the transient solution (\ref{ep12trans}). It then takes few e-folds to amplify the scalar spectrum ($\sim 6$). The remaining 22 e-folds (see Fig. (\ref{f1})) are essentially the duration of PDSR from $t_*$ until the amplification begins.\\ 
The amplitude of the scalar spectrum as a function of $N$ is plotted in  Figure (\ref{f1}) for the models I and IV. The corresponding observables are presented in  Table (\ref{tabexamples}).

\begin{table}[h!]
\caption{Models I-IV predictions}
\vspace{0.1 cm}
\centering
\begin{tabular}{|c || c |c |c |c |c|}
\hline
  & $c_{s,*}$ & $r$ & $\rd n_s/\rd \ln k$ & $\Delta N$ & $M_{\rm BH}/M_{\odot}$  \\
[0.1ex]
\hline
\hline
Model I	& 0.12	& 0.15	& 0.043	& 30 & $10^{-16}$\\
Model II	& 0.09	& 0.04	& -0.005	& 75 & $10^{-55}$\\
Model III	& 0.17	& 0.51	& 0.285	& 14 & $10^{-2}$\\
\hline
Model IV	& 0.11 	& 0.03	& -0.0003	& 28 & $10^{-14}$\\
[1ex]
\hline
\end{tabular}
\label{tabexamples}
\end{table}
\section{Conclusions}
We studied the possible consequences of a specific, non-canonical, kinetic term for the inflaton field on the spectrum of scalar perturbations generated by the inflationary expansion. The presence of a kinetic term, similar to that which is responsible for the accelerated expansion in k-inflation model, and the addition of a potential term can lead to  an inflationary phase which consists of two distinct dynamical regimes. When the potential dominates, the evolution of primordial perturbations leads to  a nearly flat spectrum. When the potential becomes negligible inflation can proceed close to another inflationary attractor. In this second phase the speed of sound associated with primordial fluctuations may be very small and thus amplify the amplitude of the scalar spectrum.\\
In particular our goal was to suggest such a mechanism for the amplification of the scalar spectrum as a source for large inhomogeneities at the end of inflation which, on horizon re-entry, could lead to the production of PBHs. To some extent the mechanism we analyse resembles that of ultra slow-roll and is based on the inverse proportionality between the amplitude of the scalar perturbations $\mathcal{P}_s$ and the product between the SR parameter $\ep{1}$ and the speed of sound $c_s$. In contrast with ultra SR, it is the varying speed of sound in our framework that can lead to a much efficient amplification of primordial inhomogeneities.\\
We finally exhibit four examples where such an  amplification does occur. These four examples are toy models chosen in order to illustrate how the suggested mechanism works, even without adopting particularly complicated potentials. The simplest model (with an exponential potential) can fit CMB data for $\alpha\sim 10^{-1}$ leading to a production of PBHs in the desired mass range, around $10^{-16}M_\odot$. Different values of $\alpha$ either generate too light PBHs or are incompatible with CMB observations. An alternative model is also studied with a slightly more complicated potential. In this latter case all the ingredients necessary in order to obtain a fast amplification of the primordial spectrum, a very good fit to the Planck data and the production of PBHs with a mass $\sim 10^{-14}M_\odot$ were introduced.\\
\section{Acknowledgements}
Alexander Y. Kamenshchik is supported in part by the RFBR grant 18-52-45016.


\end{document}